\documentclass[11pt]{article}
\usepackage{amsfonts}
\usepackage{latexsym,amsmath}
\usepackage{amssymb,array}
\makeatletter \oddsidemargin 0in \evensidemargin 0in \textwidth
16cm\textheight 20cm 
\newtheorem{r1}{Remark}[section]

\newtheorem{a1}{Assumption}[section]

\begin{document}
\title{Parameter Estimates of General Failure Rate Model:\\ A Bayesian Approach}
\author{Asok K. Nanda\footnote{Corresponding
author e-mail:
asok.k.nanda@gmail.com}\\
Department of Mathematics and Statistics\\
Indian Institute of Science Education and Research Kolkata\\
Mohanpur Campus\\ Mohanpur 741252, India. \and
Sudhansu S. Maiti\\
Department of Statistics\\ Visva-Bharati University\\
Santiniketan 731 235, India\and
Chanchal Kundu\\
Department of Mathematics\\
Rajiv Gandhi Institute of Petroleum Technology\\Rae Bareli 229 316, India\and Amarjit Kundu\\Department of
Mathematics\\Santipur College\\P.O. Santipur, Dist. Nadia\\ West
Bengal, India}
\date{September, 2013} \maketitle
\begin{abstract}The failure rate function plays an important role in
studying the lifetime distributions in reliability theory and life
testing models. A study of the general failure rate model
$r(t)=a+bt^{\theta-1},$ under squared error loss function taking
$a$ and $b$ independent exponential random variables has been
analyzed in the literature. In this article, we consider $a$ and
$b$ not necessarily independent. The estimates of the parameters
$a$ and $b$ under squared error loss, linex loss and entropy loss
functions are obtained here.
\end{abstract}
\section{Introduction}
Failure is an unavoidable phenomenon with technological products
and systems. Reliability is a measure of failure uncertainty. The
failure rate function plays an important role in reliability
theory. The failure rate function and the distribution function
are equivalent in the sense that knowing one, other can be
uniquely determined by the relationship
$$\bar{F}_X(t)=1-F_X(t)=e^{-\int_0^tr_X(u)du},$$ or
equivalently, $$f_X(t)=r_{X}(t)e^{-\int_0^tr_X(u)du},$$ where
$f_X,~F_X,~\bar{F}_X$ and $r_X$ are the probability density, the
distribution, the survival and the failure rate functions,
respectively.\\
\hspace*{.2in} In reliability, lifetime distributions are often
specified by choosing a particular failure rate function. The
constant failure rate characterizes exponential distribution. The
linear failure rate (LFR) distribution arises often in reliability
literature and is motivated by its application to human survival
data, see, for instance, Kodlin \cite{k67} and Carbone,
Kellerhouse and Gehan \cite{ckg67}. Various distributional
properties and applications of the LFR distribution to life
testing and reliability studies have been described by Sen
\cite{s06} and the references therein. Estimation of the unknown
parameters in the lifetime distributions of the individual
components belonging to a multi-component system is an interesting
problem in reliability analysis. These estimators may be extremely
useful in some ways, since they reflect the component reliability
after being assembled into an operational system (cf. Usher and Hodgson,
\cite{uh88}). Such estimators can be used under appropriate
conditions to predict the reliability of new configurations of the
components of the system. The estimation of the LFR model
$r_X(t)=a+bt$ with $a$ and $b$ nonnegative constants, using
classical method, has been studied in the literature by Bain
\cite{b74}, Shaked \cite{s78}, Sen and
Bhattacharyya \cite{sb95} among others.\\
\hspace*{.2in} Ashour and Youssef \cite{ay91} have investigated
the Bayesian estimators for the parameters of the LFR model based
on Type-II censored samples. However, their derivation of the
marginal posterior distributions seems to be erroneous (cf. Lin,
Wu and Balakrishnan \cite{lwb06}). Pandey, Singh and Zimmer
\cite{psz93} have discussed the same problem with a simpler
assumption on the joint prior distribution. The Bayesian
estimation and prediction problems for the LFR model under general
progressively Type-II censored samples are considered by Lin, Wu
and Balakrishnan \cite{lwb06}. The estimation of LFR distribution based on records and inter-record times has been discussed in Lin, Wu and Balakrishnan \cite{lwb03}. In life testing and reliability
studies, LFR distributions are useful in modeling the life length
of a system or component when failures occur at random, and also
from ageing or wear-out. But this failure rate model cannot
describe other failure rate models except the linearly increasing
one. To overcome this difficulty and to accommodate more varieties
of failure rate models, Sarhan \cite{s99} has analyzed a more
general failure rate model of the form
\begin{equation}
\label{a0}
 r_X(t)=a+bt^{\theta -1},
\end{equation}
 for some nonnegative constants
$a,b$ and $\theta$. (\ref{a0}) generalizes exponential
distribution ($b=0$ or $\theta=1$), Rayleigh distribution
($a=0,~\theta=2$), Weibull distribution ($a=0$) and LFR model
($\theta=2$). For $0\leqslant\theta\leqslant 1$, $(\ref{a0})$
gives DFR (decreasing in failure rate) distribution, whereas for
$\theta\geqslant 1$, it gives IFR (increasing in failure rate)
distributions. The general failure rate model given in (\ref{a0})
is a useful model to specify the lifetime
distributions in reliability theory and life testing.\\
\hspace*{.2in} Sarhan \cite{s99} has obtained the Bayes'
estimators of ($a$, $b$) under the squared error loss function
taking $a$ and $b$ independent exponential random variables with
known parameters, and $\theta$ a known constant. But, $a$ and
$b$ may not always be independent. Also, several common situations
may arise when over estimation is more serious than under
estimation and vice versa. In that case the loss function cannot
be symmetric (e.g., squared error loss). Here we consider the case
when $a$ and $b$ are not necessarily independent and the data are
Type-II censored. We assume that $\theta$ is known and
$(a,b)$ have joint probability density function
\begin{equation}
\label{a1} h(a,b)=f(a)g(b)+\rho
f(a)g(b)[1-2F(a)][1-2G(b)],\;\;\;-1\leqslant\rho\leqslant 1
\end{equation}
where $f$ and $g$ are the marginal probability densities of $a$
and $b$, and $F$ and $G$ are the distribution functions
corresponding to $f$ and $g,$ respectively. The probability
density function given in ($\ref{a1}$) is the well known {\it
Farlie-Gumbell-Morgenstern} bivariate density (cf. Farlie,
\cite{f60}). We assume that the value of $\rho$ in $(\ref{a1})$ is
known. Clearly, when $\rho=0$, we get the result of Sarhan
\cite{s99}. It is to be mentioned here that parameter estimation of
the general failure rate model (\ref{a0}) using masked data is
considered in Sarhan \cite{s04}.\\
\hspace*{.2in} The paper is arranged as follows. In Section 2, we
formulate the problem. The Bayes' estimators of the parameters $a$
and $b$ under different loss functions $viz.$ squared error loss
function, linex loss function and entropy loss function are
obtained in Section 3. In Section 4 some simulation results are
presented.
\section{Formulation of the problem}
 \hspace*{.2in} In the present context, we
use the following assumptions:
\begin{a1}
\label{as1}
A fixed number of units, say $n$, are put on test and
the data collected on the life of the units are Type-II censored.
That is, the test is terminated once $r$ (prespecified number)
units fail. The failure times of the first $r$ units are known.
Let they be $t_1,t_2,\ldots,t_r$.
\end{a1}
\begin{a1}
\label{as2}
No two units fail together. This means, $t_1<t_2<\ldots<t_r$.
\end{a1}
\begin{a1}\label{as3}
The marginal distributions of $a$ and $b$ are exponential with
known means $1/\lambda_1$ and $1/\lambda_2,$ respectively.
\end{a1}
\begin{a1}\label{as4}
Failure times of the units are statistically independent.
\end{a1}
\hspace*{.2in} Under Assumption $\ref{as3}$, $(\ref{a1})$ becomes
$$h(a,b)=\lambda_1\lambda_2e^{-\lambda_1a-\lambda_2b}+\lambda_1
\lambda_2\rho\left(2e^{-2\lambda_1a}-e^{-\lambda_1a}\right)
\left(2e^{-2\lambda_2b}-e^{-\lambda_2b}\right).$$ Given $a$ and
$b$, the survival function corresponding to $(\ref{a0})$ is given
by
\begin{equation}
\label{a2}
 \bar{F}_X(t|a,b)=\exp\left[-\left(at+bt^\theta/\theta\right)\right],~ t\geqslant 0.
\end{equation}
\hspace*{.2in} If $T$ denotes the random variable having failure
rate function given by $(\ref{a0})$, then the joint probability
density function of $(a,b)$ and $T$ is given by
\begin{eqnarray*}
g_1(t,a,b)&=&\left(a+bt^{\theta-1}\right)\exp\left[-\left(at+bt^\theta/\theta\right)\right]
\left[\lambda_1\lambda_2e^{-\lambda_1a-\lambda_2b}\right.\\&&\left.+\lambda_1
\lambda_2\rho\left(2e^{-2\lambda_1a}-e^{-\lambda_1a}\right)
\left(2e^{-2\lambda_2b}-e^{-\lambda_2b}\right)\right],
\end{eqnarray*}
 which gives the
marginal probability density function of $T$, after some
simplifications, as
\begin{eqnarray*}\label{u1}
f_T(t)&=&\int_0^\infty\int_0^\infty g_1(t,a,b)da\;db\\
&=&\lambda_1\lambda_2\int_0^\infty\int_0^\infty(a+bt^{\theta-1})e^{-(at+bt^\theta/\theta)}
e^{-\lambda_1a-\lambda_2b}da\;db\\
&&+\lambda_1\lambda_2\rho\left[4\int_0^\infty\int_0^\infty
(a+bt^{\theta-1})e^{-(at+bt^\theta/\theta)}e^{-2\lambda_1a-2\lambda_2b}da\;db\right.\\
&&-2\int_0^\infty\int_0^\infty(a+bt^{\theta-1})e^{-(at+bt^\theta/\theta)}e^{-\lambda_1a-2\lambda_2b}da\;db\\
&&-2\int_0^\infty\int_0^\infty(a+bt^{\theta-1})e^{-(at+bt^\theta/\theta)}
e^{-2\lambda_1a-\lambda_2b}da\;db\\
&&\left.+\int_0^\infty\int_0^\infty
(a+bt^{\theta-1})e^{-(at+bt^\theta/\theta)}
e^{-\lambda_1a-\lambda_2b}da\;db\right]\\
&=&\lambda_1\lambda_2I_1+\lambda_1\lambda_2\rho\left[4I_2-2I_3-2I_4+I_1\right],~
{\rm say},
\end{eqnarray*}
where
\begin{eqnarray*}
I_1&=&\int_0^\infty\int_0^\infty(a+bt^{\theta-1})e^{-(at+bt^\theta/\theta)}
e^{-\lambda_1a-\lambda_2b}da\;db\\
&=&\frac{\theta}{(t+\lambda_1)^2(t^\theta+\theta\lambda_2)}+
\frac{\theta^2t^{\theta-1}}{(t+\lambda_1)(t^\theta+\theta\lambda_2)^2},\\
I_2&=&\int_0^\infty\int_0^\infty
(a+bt^{\theta-1})e^{-(at+bt^\theta/\theta)}e^{-2\lambda_1a-2\lambda_2b}da\;db\\
&=&\frac{\theta}{(t+2\lambda_1)^2(t^\theta+2\theta\lambda_2)}+
\frac{\theta^2t^{\theta-1}}{(t+2\lambda_1)(t^\theta+2\theta\lambda_2)^2},\\
I_3&=&\int_0^\infty\int_0^\infty(a+bt^{\theta-1})e^{-(at+bt^\theta/\theta)}e^{-\lambda_1a-2\lambda_2b}da\;db\\
&=&\frac{\theta}{(t+\lambda_1)^2(t^\theta+2\theta\lambda_2)}+
\frac{\theta^2t^{\theta-1}}{(t+\lambda_1)(t^\theta+2\theta\lambda_2)^2}
\end{eqnarray*}and
\begin{eqnarray*}
I_4&=&\int_0^\infty\int_0^\infty(a+bt^{\theta-1})e^{-(at+bt^\theta/\theta)}
e^{-2\lambda_1a-\lambda_2b}da\;db\\
&=&\frac{\theta}{(t+2\lambda_1)^2(t^\theta+\theta\lambda_2)}+
\frac{\theta^2t^{\theta-1}}{(t+2\lambda_1)(t^\theta+\theta\lambda_2)^2}.
\end{eqnarray*}
Thus, we have
\begin{eqnarray}
f_T(t)&=&\frac{\lambda_1\lambda_2\theta\left\{(\theta+1)t^\theta+\theta
\lambda_1t^{\theta-1}+ \theta\lambda_2\right\}}
{\left(\lambda_1+t\right)^2\left(t^\theta+\theta\lambda_2\right)^2}
\nonumber\\&& +\frac{\rho\lambda_1\lambda_2\theta t^\theta}
{\left(2\lambda_2\theta+t^\theta\right)\left(\lambda_2\theta+t^\theta
\right)\left
(2\lambda_1+t\right)\left(\lambda_1+t\right)}\nonumber\\&&
\left[\frac{t^2-2\lambda_1^2}{\left(2\lambda_1+t\right)\left
(\lambda_1+t\right)}
+\frac{\theta\left(t^{2\theta}-2\lambda_2^2\theta^2\right)}{\left(2
\lambda_2\theta+t^\theta\right)
\left(\lambda_2\theta+t^\theta\right)}\right]\label{a3}.
\end{eqnarray}
\begin{r1}
If it is assumed that $\rho$ is unknown with a uniform prior
distribution in ($-1, 1$), then the probability density function of $T$ becomes
$f_T(t)=\lambda_1\lambda_2I_1,$
which is same as ($\ref{a3}$) with $\rho=0$.
\end{r1}
\section{Bayes' estimators of $a$ and $b$}
Once the values of $t_1,t_2,\ldots,t_r$ and that of $n$ are known,
the likelihood function can be written as (cf. Lawless \cite{l})
\begin{equation}
\label{a4} L({\bf
t}|a,b)=\Pi_{i=1}^rf_X(t_i|a,b)\left[\bar{F}_X(t_r|a,b)\right]^{n-r},
\end{equation}
where ${\bf t}=(t_1,t_2,\ldots t_r)$. On using $(\ref{a2})$,
$(\ref{a4})$ reduces to
\begin{eqnarray*}
L({\bf t}|a,b)
&=&\left[\Pi_{i=1}^r\left(a+bt_i^{\theta-1}\right)\right] e^{-aS_1
-\frac{b}{\theta}S_2}\\
&=&\left[\sum_{j=0}^ra^{r-j}b^jM_j({\bf t})\right] e^{-aS_1-
\frac{b}{\theta}S_2},
\end{eqnarray*}
where
\begin{eqnarray*}
S_1&=&\sum_{i=1}^rt_i+(n-r)t_r,\\
S_2&=&\sum_{i=1}^rt_i^\theta+(n-r)t_r^\theta,
\end{eqnarray*}
and
\begin{eqnarray*}
 M_j({\bf t})&=&\sum_{1\leqslant i_1\leqslant\ldots}\ldots\sum_{\leqslant i_j\leqslant
r}t_{i_1}^{\theta-1}t_{i_2}^{\theta-1}\ldots t_{i_j}^{\theta-1},
\end{eqnarray*}
for $j=1,2,\ldots,r$ with $M_0({\bf t})=1$. A similar kind of expression may be obtained in Sarhan \cite{s99}. Now, the joint
distribution of ${\bf t},~a$ and $b$ is
\begin{eqnarray*}
f^*(a,b,{\bf t})&=&L({\bf t}|a,b)h(a,b)\\
&=&\sum_{j=0}^ra^{r-j}b^jM_j({\bf
t})e^{-aS_1-\frac{b}{\theta}S_2}\left[\lambda_1\lambda_2e^{-\lambda_1a-\lambda_2b}\right.\\
&&+\left.\lambda_1\lambda_2\rho\left\{4e^{-2\lambda_1a-2\lambda_2b}-2e^{-\lambda_1a-2\lambda_2b}
-2e^{-2\lambda_1a-\lambda_2b}+e^{-\lambda_1a-\lambda_2b}\right\}\right]\\
&=&\lambda_1\lambda_2\sum_{j=0}^ra^{r-j}b^jM_j({\bf
t})\left[e^{-a(S_1+\lambda_1)-b(\frac{S_2}{\theta}+\lambda_2)}\right.\\
&&+\rho\left\{4e^{-a(S_1+2\lambda_1)-b(\frac{S_2}{\theta}+2\lambda_2)}
-2e^{-a(S_1+\lambda_1)-b(\frac{S_2}{\theta}+2\lambda_2)}\right.\\
&&-\left.\left.2e^{-a(S_1+2\lambda_1)-b(\frac{S_2}{\theta}+\lambda_2)}
+e^{-a(S_1+\lambda_1)-b(\frac{S_2}{\theta}+\lambda_2)}\right\}\right],
\end{eqnarray*}
which gives the marginal density function of ${\bf t}$ as
\begin{eqnarray*}
f_1^*({\bf t})&=&\int_0^\infty\int_0^\infty f^*(a,b,{\bf t})da\;db\\
&=&\int_0^\infty\int_0^\infty
\lambda_1\lambda_2\sum_{j=0}^ra^{r-j}b^jM_j({\bf
t})\left[e^{-a(S_1+\lambda_1)-b(\frac{S_2}{\theta}+\lambda_2)}\right.\\
&&+\rho\left\{4e^{-a(S_1+2\lambda_1)-b(\frac{S_2}{\theta}+2\lambda_2)}
-2e^{-a(S_1+\lambda_1)-b(\frac{S_2}{\theta}+2\lambda_2)}\right.\\
&&-\left.\left.2e^{-a(S_1+2\lambda_1)-b(\frac{S_2}{\theta}+\lambda_2)}
+e^{-a(S_1+\lambda_1)-b(\frac{S_2}{\theta}+\lambda_2)}\right\}\right]da~db\\
&=&\lambda_1\lambda_2\left[I_5+\rho\left\{4I_6-2I_7-2I_8+I_5\right\}\right],~{\rm
say},
\end{eqnarray*}
where
\begin{eqnarray*}
I_5&=&\int_0^\infty\int_0^\infty\sum_{j=0}^ra^{r-j}b^jM_j({\bf
t})e^{-a(S_1+\lambda_1)-b(\frac{S_2}{\theta}+\lambda_2)}da\;db\\
&=&\sum_{j=0}^rM_j({\bf t})\int_0^\infty
a^{r-j}e^{-a(S_1+\lambda_1)}da\int_0^\infty
b^je^{-b(\frac{S_2}{\theta}+\lambda_2)}db\\
&=&\sum_{j=0}^rM_j({\bf t})\frac{\Gamma
(r-j+1)}{(S_1+\lambda_1)^{r-j+1}}\frac{\Gamma
(j+1)}{(\frac{S_2}{\theta}+\lambda_2)^{j+1}},\\
&=&\Phi(1,1,1,1),
\end{eqnarray*}
\begin{eqnarray*}
I_6&=&\int_0^\infty\int_0^\infty\sum_{j=0}^ra^{r-j}b^jM_j({\bf
t})e^{-a(S_1+2\lambda_1)-b(\frac{S_2}{\theta}+2\lambda_2)}da\;db\\
&=&\sum_{j=0}^rM_j({\bf t})\frac{\Gamma(
r-j+1)}{(S_1+2\lambda_1)^{r-j+1}}\frac{\Gamma
(j+1)}{(\frac{S_2}{\theta}+2\lambda_2)^{j+1}},\\
&=&\Phi(1,1,2,2),
\end{eqnarray*}
\begin{eqnarray*}
I_7&=&\int_0^\infty\int_0^\infty\sum_{j=0}^ra^{r-j}b^jM_j({\bf
t})e^{-a(S_1+\lambda_1)-b(\frac{S_2}{\theta}+2\lambda_2)}da\;db\\
&=&\sum_{j=0}^rM_j({\bf t})\frac{\Gamma
(r-j+1)}{(S_1+\lambda_1)^{r-j+1}}\frac{\Gamma
(j+1)}{(\frac{S_2}{\theta}+2\lambda_2)^{j+1}}\\
&=&\Phi(1,1,1,2)
\end{eqnarray*}and
\begin{eqnarray*}
I_8&=&\int_0^\infty\int_0^\infty\sum_{j=0}^ra^{r-j}b^jM_j({\bf
t})e^{-a(S_1+2\lambda_1)-b(\frac{S_2}{\theta}+\lambda_2)}da\;db\\
&=&\sum_{j=0}^rM_j({\bf t})\frac{\Gamma
(r-j+1)}{(S_1+2\lambda_1)^{r-j+1}}\frac{\Gamma
(j+1)}{(\frac{S_2}{\theta}+\lambda_2)^{j+1}},\\
&=&\Phi(1,1,2,1)
\end{eqnarray*}
so that
\begin{eqnarray*}
 f_1^*({\bf t})&=&\lambda_1\lambda_2\left[\Phi(1,1,1,1)+\rho\left\{4\Phi(1,1,2,2)
-2\Phi(1,1,1,2)-2\Phi(1,1,2,1)+\Phi(1,1,1,1)\right\}\right]\\
&=&K^{-1},~{\rm say},
\end{eqnarray*}
where
$$\Phi(l,m,p,q)=\sum_{j=0}^rM_j({\bf t})
\frac{\Gamma{(r-j+l)}}{a_{1p}^{r-j+l}}.\frac{\Gamma{(j+m)}}
{a_{2q}^{j+m}},$$
$a_{1p}=S_1+p\lambda_1$ and $a_{2q}=\frac{S_2}{\theta}+q\lambda_2.$ \\
\hspace*{.2in} Hence the posterior joint probability density
function of $(a,b)$ is given by
\begin{eqnarray*}
f_2(a,b|{\bf t})&=&\frac{f^*(a,b,{\bf t})}{f_1^*({\bf t})}\\
&=&K\lambda_1\lambda_2\sum_{j=0}^ra^{r-j}b^jM_j({\bf
t})\left[e^{-a(S_1+\lambda_1)-b(\frac{S_2}{\theta}+\lambda_2)}\right.\\
&&+\rho\left\{4e^{-a(S_1+2\lambda_1)-b(\frac{S_2}{\theta}+2\lambda_2)}
-2e^{-a(S_1+\lambda_1)-b(\frac{S_2}{\theta}+2\lambda_2)}\right.\\
&&-\left.\left.2e^{-a(S_1+2\lambda_1)-b(\frac{S_2}{\theta}+\lambda_2)}
+e^{-a(S_1+\lambda_1)-b(\frac{S_2}{\theta}+\lambda_2)}\right\}\right].
\end{eqnarray*}
\subsection{Estimation under squared error loss function}
\hspace*{.2in} Let us consider the loss function of the form
\begin{equation}
\label{a5}
L_1^*(\psi,\hat{\psi})=k_1(a-\hat{a})^2+k_2(b-\hat{b})^2,
\end{equation}
where $k_1,~k_2>0,\;\psi=(a,b)$, and
$\hat{\psi}=(\hat{a},\hat{b})$ is the estimator of $\psi$. It is
well known that under the loss function of the form $(\ref{a5})$,
Bayes' estimator of a parameter is its posterior mean. Thus,
Bayes' estimators of $a$ and $b$ are given by
$$\hat{a}_{BS}=E_{f_2}(a)=\int_0^\infty\int_0^\infty
af_2(a,b|{\bf t})da\;db$$ and
$$\hat{b}_{BS}=E_{f_2}(b)=\int_0^\infty\int_0^\infty bf_2(a,b|{\bf t})da\;db.$$
These, after simplification, reduce respectively to
$$\hat{a}_{BS}=K\lambda_1\lambda_2[\Phi(2,1,1,1)+\rho\{4\Phi(2,1,2,2)-2\Phi(2,1,1,2)
-2\Phi(2,1,2,1)+\Phi(2,1,1,1)\}]$$ and
$$\hat{b}_{BS}=K\lambda_1\lambda_2[\Phi(1,2,1,1)+\rho\{4\Phi(1,2,2,2)-2\Phi(1,2,1,2)
-2\Phi(1,2,2,1)+\Phi(1,2,1,1)\}].$$
\subsection{Estimation under linex loss function}
Here we consider the loss function of the form
$$
L_2^*(\psi,\hat{\psi})=l_1\left[e^{c_1(\hat{a}-a)}-c_1(\hat{a}-a)-1\right]+l_2\left[e^{c_2(\hat{b}-b)}-c_2(\hat{b}-b)-1\right],
$$
where $c_1,~c_2,~l_1,~l_2$ are constants, and $\psi$ and
$\hat{\psi}$
are as defined earlier.\\
\hspace*{.2in} The Bayes' estimators of $a$ and $b$ are then given
by
\begin{eqnarray*}
\hat{a}_{BL}&=&-\frac{1}{c_1}\ln E_{f_2}\left(e^{-c_1a}|{\bf
t}\right)\\&=& -\frac{1}{c_1}\ln \left[\int_0^\infty\int_0^\infty
e^{-c_1a}f_2(a,b|{\bf t})da~db\right]
\end{eqnarray*}
and
\begin{eqnarray*}
\hat{b}_{BL}&=&-\frac{1}{c_2}\ln E_{f_2}\left(e^{-c_2b}|{\bf
t}\right)\\&=& -\frac{1}{c_2}\ln \left[\int_0^\infty\int_0^\infty
e^{-c_2b}f_2(a,b|{\bf t})da~db\right],
\end{eqnarray*}
respectively. These, after simplification, reduce respectively to
{\small$$\hat{a}_{BL}=-\frac{1}{c_1}\ln\left\{K\lambda_1\lambda_2[\Phi^*(1,1,1,1)+\rho\{4\Phi^*(1,1,2,2)-2\Phi^*(1,1,2,1)
-2\Phi^*(1,1,1,2)+\Phi^*(1,1,1,1)\}]\right\}$$} and
{\small$$\hat{b}_{BL}=-\frac{1}{c_2}\ln\left\{K\lambda_1\lambda_2[\Phi^{**}(1,1,1,1)+\rho\{4\Phi^{**}(1,1,2,2)-2\Phi^{**}(1,1,2,1)
-2\Phi^{**}(1,1,1,2)+\Phi^{**}(1,1,1,1)\}]\right\},$$} where
$\Phi^*=\Phi$ with $a_{1p}$ replaced by
$a^*_{1p}=c_1+S_1+p\lambda_1$ and $\Phi^{**}=\Phi$ with $a_{2q}$
replaced by $a^*_{2q}=c_2+\frac{S_2}{\theta}+q\lambda_2$.
\subsection{Estimation under entropy loss function}
Here we consider the loss function of the form
$$
L_3^*(\psi,\hat{\psi})=m_1\left[\frac{\hat{a}}{a}-\ln
\frac{\hat{a}}{a}-1\right]+m_2\left[\frac{\hat{b}}{b}-\ln
\frac{\hat {b}}{b}-1\right],
$$
where $m_1$ and $m_2$ are constants, and $\psi$ and $\hat{\psi}$
are as defined earlier.\\
\hspace*{.2in} The Bayes' estimators of $a$ and $b$ are then given
by
\begin{eqnarray*}
\hat a_{BE}=\frac{1}{E_{f_2}\left(\frac{1}{a}|{\bf t}\right)}
\end{eqnarray*}
and
\begin{eqnarray*}
\hat b_{BE}=\frac{1}{E_{f_2}\left(\frac{1}{b}|{\bf t}\right)}
\end{eqnarray*}
respectively, which, after simplification, reduce respectively to
$$\hat a_{BE}=\left[K\lambda_1\lambda_2\left\{\Phi(0,1,1,1)+\rho(4\Phi(0,1,2,2)-2\Phi(0,1,2,1)
-2\Phi(0,1,1,2)+\Phi(0,1,1,1))\right\}\right]^{-1}$$ and
$$\hat b_{BE}=\left[K\lambda_1\lambda_2\left\{\Phi(1,0,1,1)+\rho(4\Phi(1,0,2,2)-2\Phi(1,0,2,1)
-2\Phi(1,0,1,2)+\Phi(1,0,1,1))\right\}\right]^{-1}.$$
\section{Simulation and Conclusion}
 Once the Bayes' estimators of the parameters $a$ and $b$ are obtained it might be of interest to know how
 the estimators behave for different values of the parameters of
 the underlying model. This is done in this section through a
 simulation study as is detailed below.\\
 \hspace*{.2in}In the tables given in the Appendix, the values of $\hat{a}_{BS}$, $\hat{b}_{BS},$ $\hat{a}_{BL},$ $\hat{b}_{BL},$
 $\hat{a}_{BE}$ and $\hat{b}_{BE}$ are given for different values of
 $n,~r,~\theta,~\lambda_1,~\lambda_2,~\rho,~c_1$ and $c_2$.
 Keeping the other parameters fixed, the values of $\hat{a}_{BS},$ $\hat{b}_{BS},$
$\hat{a}_{BL},$ $\hat{b}_{BL},$ $\hat{a}_{BE}$ and $\hat{b}_{BE},$
for different values of $n$, $r$, $\theta$, $\rho$, $\lambda_1$
and $\lambda_2,$ are given in Tables 2-7 respectively. Note that here we have simulated more number of values to get a clear picture of $\hat{a}_{BE}$. Those values are reported at the bottom portion of Table 3. It is to be
mentioned here that $\hat{a}_{BS},$ $\hat{b}_{BS},$ $\hat{a}_{BE}$
and $\hat{b}_{BE}$ are independent of $c_1$ and $c_2$, whereas
$\hat{a}_{BL}$ and $\hat{b}_{BL}$ vary with $c_1$ and $c_2$ respectively.\\
\hspace*{.2in} Table $8$ shows the values of $\hat{a}_{BL}$ and
$\hat{b}_{BL}$ for different values of $c_1$ and $c_2$
respectively, taking the other parameters fixed. In Table $1$ given below the
conclusions about the monotonicity of $\hat{a}_{BS},$
$\hat{b}_{BS},$ $\hat{a}_{BL},$ $\hat{b}_{BL},$ $\hat{a}_{BE}$ and
$\hat{b}_{BE}$ have been made on the basis of the tabulated values. Once the values show non-monotone behavior, we have simulated for more number of values, and the specific non-monotonic behavior has been noted in Table $1$.\\
\hspace*{.2in} To compute these values following methodology has
been adopted and the simulation works have been done using R-Software.
\begin{enumerate}
\item[($1$)]Markov Chain Monte Carlo (MCMC) method (using
Metropolis-Hastings algorithm) has been adopted to generate
samples from $f_T(t),$ the target distribution, given in
($\ref{u1}$). We take proposal distribution as exponential with
rate $(\lambda_1+\lambda_2)/2$. In the simulation process, we take
$5000$ burn-in observations. Then, a sample of size $n$ is chosen
and out of these observations, Type-II censored sample
$(t_1,t_2,\ldots,t_r)$ is collected censoring at the point $r$.
$1000$ such repeated samples are collected and required
calculations made. Between two consecutive samples, 100
observations are discarded to minimize the dependency.
\item[($2$)]For $j=1,2,\ldots, r$, the values of $M_j({\bf t})$
are calculated from the values of $t_1,t_2,\ldots,t_r$.
\item[($3$)]Taking some fixed values of
$n,~r,~\theta,~\lambda_1,~\lambda_2,~\rho,~c_1$ and $c_2$, the
values of $\Phi(l,m,p,q)$, $\Phi^*(l,m,p,q)$ and
$\Phi^{**}(l,m,p,q)$ are calculated for different values of
$l,\;m,\;p$ and $q$, which give the values of $\hat{a}_{BS}$,
$\hat{b}_{BS},$ $\hat{a}_{BL},$ $\hat{b}_{BL},$
 $\hat{a}_{BE}$ and $\hat{b}_{BE}$.
\item[($4$)]Averages are calculated from these $1000$ values
of $\hat{a}_{BS}$, $\hat{b}_{BS},$ $\hat{a}_{BL},$ $\hat{b}_{BL},$
 $\hat{a}_{BE}$ and $\hat{b}_{BE}$, which give their final
 values. Variances of these $\hat{a}_{BS}$,
$\hat{b}_{BS},$ $\hat{a}_{BL},$ $\hat{b}_{BL},$
 $\hat{a}_{BE}$ and $\hat{b}_{BE}$ are also calculated and it is
 clear from Table $9$ that all are decreasing functions of $n$, as expected.
 \item[($5$)]Empirical Bayes risk for estimating $a$ and $b$ under different
loss functions are also calculated using the definition
$$R^E_i=\frac{1}{1000}\sum_{j=1}^{1000}L^{*}_i(\bar\psi,\hat\psi_j),~\mbox{where}~\bar\psi=\frac{1}{1000}\sum_{j=1}^{1000}\hat\psi_j,$$
$i=1,2,3$. Here $i=1$ gives $R_{BS}^E$, $i=2$ gives $R_{BL}^E$, and $i=3$ gives $R_{BE}^E$. Accordingly we define $R^E_{BS}$, $R^E_{BL}$ and $R^E_{BE}$ as empirical Bayes' risk under
squared error loss function, linex loss function and entropy loss function respectively. Without any loss of generality, we choose each of $k_1,\;k_2,\;l_1,\;l_2$ and $m_1,\;m_2$ as unity. Taking fixed values of
$n,~r,~\theta,~\lambda_1,~\lambda_2,~c_1$ and $c_2$, as in Table 4 the
values of $R^E_{BS}$, $R^E_{BL}$ and $R^E_{BE}$ are obtained under 1000 repetitions for different values of $\rho$. It is observed from Table 10 that $R^E_{BS}$ is decreasing in $\rho$. It is observed from Table 10 that the values of $R^E_{BS}$ is less for any $\rho>0$ than that when $a$ and $b$
are taken as independent, i.e., $\rho=0$. Similarly, we can find some $\rho$ for which $R^E_{BL}$ and $R^E_{BE}$ are less
than the risk for $\rho=0$, which exhibits the necessity of taking $a$ and $b$ not independent.
\end{enumerate}
We have chosen basically two types of loss
functions - symmetric (squared error loss) and asymmetric (linex
loss and entropy loss). Two asymmetric loss functions have been
chosen for their distinctive features - linex loss function has
different shapes depending on $c_1$ and $c_2$, whereas entropy
loss function has no change in shape. It may be possible to choose
a number of loss functions but these three are chosen as
representatives.\\
\hspace*{.2in}Once the estimates of $a$ and $b$ are obtained, the
estimates of the reliability function and the failure rate
function can be computed using the relationships
\begin{eqnarray*}
\hat{\bar{F}}_X(t)=\exp\left[-\left(\hat at+\hat
bt^\theta/\theta\right)\right]
\end{eqnarray*}
and
\begin{eqnarray*}
\hat r(t)=\hat a+\hat bt^{\theta-1}.
\end{eqnarray*}
For different values of $t$, the estimates of these two
reliability measures can be calculated.
\newpage

\begin{table}[h]
\caption{Conclusions}
\begin{center}
\begin{tabular}{|c|c|c|c|c|c|c|}\hline
${\rm parameters}$ &$\hat{a}_{BS}$ &$\hat{b}_{BS}$ &$\hat{a}_{BL}$
&$\hat{b}_{BL}$&$\hat{a}_{BE}$&$\hat{b}_{BE}$\\
\hline$\;n\;$ & RC  & increasing & increasing  & increasing &
RC  & increasing\\
\hline$\;r\;$ & decreasing & decreasing & BT & decreasing
&RC & decreasing \\
\hline$\;\theta\;$ & RC & increasing & UBT & BT & RC &
increasing\\
\hline$\;\rho\;$ & decreasing & decreasing & RC & decreasing &
RC & decreasing\\
\hline$\;\lambda_1\;$ & decreasing & increasing & decreasing &
increasing& decreasing & BT\\
\hline$\lambda_2$ & BT & decreasing & increasing & decreasing & BT & decreasing\\
\hline$c_1$ & unchanged & unchanged & decreasing & unchanged &
unchanged& unchanged\\
\hline$c_2$ &unchanged & unchanged & unchanged & decreasing &
unchanged & unchanged\\\hline
\end{tabular}
\end{center}
\end{table}
The following abbreviations have been used in the above table.\\

BT: Bathtub shaped; \\
UBT: Upside down bathtub shaped;\\
RC: Roller coaster.

\newpage
\begin{center}
{\bf \Large Appendix\footnote{Some additional simulation results in support of Table 1
for estimates which show non-monotone behavior are placed on the website: ftp://210.212.53.189:4777 (user id: ftpadmin, password: rgipt@7890) }}
\end{center}
\begin{table}[h]
\caption{Table for different values of $n$
($r=15,~\theta=1.5,~\lambda_1=0.1,~\lambda_2=0.2,~\rho=0.5,~c_1=5,$
$c_2=10$).}
\begin{center}
\begin{tabular}{|c|c|c|c|c|c|c|}\hline $n$ &$\hat{a}_{BS}$
&$\hat{b}_{BS}$ &$\hat{a}_{BL}$ &$\hat{b}_{BL}$ &$\hat{a}_{BE}$
&$\hat{b}_{BE}$\\
\hline$\;\;20\;\;$ & $2.428505$ & $2.470713$ & $0.9571143$ &
$0.3315961$ & $1.841096$ & $ 1.108504$\\
\hline $\;\;30\;\;$ & $2.47891$ & $2.809811$ & $0.9602245$&
$0.3504222$ & $1.869938$ & $1.24303$\\
\hline$\;\;50\;\;$ & $2.570068$ & $3.526479$ & $1.031021$ &
$0.3728826$ & $1.981042$ & $1.544343$\\
\hline $\;\;80\;\;$ & $2.702363$ & $3.833807$ & $1.106235$&
$0.3769399$ & $2.144507$ & $1.685688$\\
\hline$\;\;90\;\;$ & $2.651358$ & $3.922087$ & $1.117268$ &
$0.3771751$ & $2.123916$ & $1.721286$\\
\hline$\;\;100\;\;$ & $2.668780$ & $3.99165$ & $1.149213$ &
$0.3780229$ & $2.161693$ & $1.764691$\\\hline
\end{tabular}
\end{center}
\end{table}

\begin{table}[h]
\caption{Table for different values of $r$
($n=50,~\theta=1.5,~\lambda_1=0.1,~\lambda_2=0.2,~\rho=0.5,~c_1=5,$
$c_2=10$).}
\begin{center}
\begin{tabular}{|c|c|c|c|c|c|}\hline $r$ &$\hat{a}_{BS}$
&$\hat{b}_{BS}$ &$\hat{a}_{BL}$ &$\hat{b}_{BL}$
&$\hat{b}_{BE}$\\
\hline$\;\;10\;\;$ & $2.889158$ & $3.872732$ & $1.046990$ &
$0.3765202$  & $1.92015$\\
\hline $\;\;15\;\;$ & $2.588836$ & $3.444316 $ & $1.045380$ &
$0.3690954$ &  $1.511899$\\
\hline$\;\;20\;\;$ & $2.476374$ & $3.041744$ & $1.073117$ &
$0.3585487$ & $1.254549$\\
\hline $\;\;25\;\;$ & $2.303392$ & $2.494415$ & $1.114468$&
$0.3398418$ &  $1.003384$\\
\hline$\;\;35\;\;$ & $2.153625$ & $1.575009$ & $1.260905$ &
$0.2871969$ &  $0.6200885$\\
\hline$\;\;45\;\;$ & $2.056847$ & $0.8700868$ & $1.416823$ &
$0.2131185$ &  $0.3430589$\\\hline
\hline
$r$ &	5&	10&	15&	20&	25\\
\hline
$\hat{a}_{BE}$& 2.767692&	2.413989	&2.11847&	2.017851&	2.010617\\
\hline
$r$&	30&	35&	40&	45&	50\\
\hline
$\hat{a}_{BE}$&	1.963753&	1.991297&	1.970913&	2.041148&	2.019716\\
\hline
\hline
$r$ &55&	60&	65&	70&	75\\
\hline
$\hat{a}_{BE}$&	2.00211	&2.005149	&2.060117&	2.058064&	2.024053\\
\hline
$r$&	80&	85&	90&	95&	100\\
\hline
$\hat{a}_{BE}$&	2.047208&	1.992579&	1.953117&	1.900956&	1.787256\\
\hline
\end{tabular}
\end{center}
\end{table}

\begin{table}[h]
\caption{Table for different values of $\theta$
($n=50,~r=15,~\lambda_1=0.1,~\lambda_2=0.2,~\rho=0.5,~c_1=5,$
$c_2=10$).}
\begin{center}
\begin{tabular}{|c|c|c|c|c|c|c|}\hline $\theta$ &$\hat{a}_{BS}$
&$\hat{b}_{BS}$ &$\hat{a}_{BL}$ &$\hat{b}_{BL}$ &$\hat{a}_{BE}$
&$\hat{b}_{BE}$\\
\hline$\;\;1.2\;\;$ & $2.53878$ & $2.698973$ & $0.7468843$ &
$0.3592404$ & $1.484473$ & $1.118551$\\
\hline $\;\;1.3\;\;$ & $2.498865$ & $2.978176$ & $0.8089464$&
$0.3638382$ & $1.619551$ & $1.247953$\\
\hline $\;\;1.5\;\;$ & $2.612741$ & $3.455287$ & $1.043905$&
$0.3682661$ & $2.023602$ & $1.511676$\\
\hline$\;\;1.7\;\;$ & $2.655687$ & $3.853542$ & $1.295282$ &
$0.3755059$ & $2.250713$ & $1.815087$\\
\hline $\;\;2.0\;\;$ & $2.617994$ & $4.128254$ & $1.586975$&
$0.3744949$ & $2.357008$ & $2.203706$\\
\hline$\;\;2.5\;\;$ & $2.649079$ & $4.335498$ & $1.803262$ &
$0.3745662$ & $2.548854$ & $2.838067$\\
\hline$\;\;3.0\;\;$ & $2.569748$ & $4.389099$ & $1.803354$ &
$0.3730589$ & $2.400356$ & $3.306292$\\
\hline$\;\;3.5\;\;$ & $2.531871$ & $4.358319$ & $1.802254$&
$0.3713863$ & $2.370501$ & $3.552629$\\
\hline$\;\;4.0\;\;$ & $2.546643$ & $4.336425$ & $1.814499$ &
$0.3706115$ & $2.386205$ & $3.736094$\\
\hline$\;\;4.5\;\;$ & $2.504744$ & $4.322051$ & $1.792483$ &
$0.370061$& $2.347539$ & $3.79589$\\
\hline$\;\;5.0\;\;$ & $2.474104$ & $4.302419$ & $1.776677$ &
$0.3694796$ & $2.319187$ & $3.846737$\\
\hline$\;\;7.0\;\;$ & $2.445197$ & $4.281886$ & $1.761036$&
$0.3689414$ & $2.292343$ & $3.894736$\\
\hline$\;\;10.0\;\;$ & $2.466696$ & $4.282187$ & $1.775499$ &
$0.3689829$ & $2.312524$ & $$3.904676\\\hline
\end{tabular}
\end{center}
\end{table}

\begin{table}[h]
\caption{Table for different values of $\rho$
($n=50,~r=15,~\theta=1.5,~\lambda_1=0.1,~\lambda_2=0.2,~c_1=5,$
$c_2=10$).}
\begin{center}
\begin{tabular}{|c|c|c|c|c|c|c|c|}\hline $\rho$ &$\hat{a}_{BS}$
&$\hat{b}_{BS}$ &$\hat{a}_{BL}$ &$\hat{b}_{BL}$ &$\hat{a}_{BE}$
&$\hat{b}_{BE}$\\
\hline$\;\;-1.0\;\;$ & $3.082916$ & $5.472054$ & $1.067603$ &
$0.4478139$ & $3.009227$ & $2.636043$\\
\hline $\;\;-0.8\;\;$ & $2.993178$ & $5.280714$ & $1.038365$&
$0.4372742$ & $2.833177$ & $2.469603$\\
\hline$\;\;-0.5\;\;$ & $2.877378$ & $4.904747$ & $1.039639$&
$0.4222705$ & $2.763539$ & $2.234952$\\
\hline $\;\;-0.1\;\;$ & $2.747384$ & $4.355944$ & $1.021720$&
$0.4007587$ & $2.598687$ & $1.927057$\\
\hline$\;\;0.0\;\;$ & $2.707445$ & $4.114485$ & $1.057667$ &
$0.3925779$ & $2.736950$ & $1.81324$\\
\hline$\;\;0.1\;\;$ & $2.687283$ & $3.867259$ & $1.045536$ &
$0.3841437$ & $2.6258$ & $1.699069$\\
\hline$\;\;0.2\;\;$ & $2.680645$ & $3.844324$ & $1.056197$ &
$0.3826008$ & $2.664627$ & $1.689498$\\
\hline$\;\;0.5\;\;$ & $2.675599$ & $3.460527$ & $1.073620$&
$0.3685548$ & $2.652695$ & $1.527531$\\
\hline$\;\;0.8\;\;$ & $2.556038$ & $3.089788$ & $1.117351$ &
$0.3578624$ & $2.623406$ & $1.379521$\\
\hline$\;\;1.0\;\;$ & $2.53029$ & $2.902948$ & $1.166945$ &
$0.3519428$ & $2.651116$ & $1.306628$\\\hline
\end{tabular}
\end{center}
\end{table}

\begin{table}[h]
\caption{Table for different values of $\lambda_1$
($n=50,~r=15,~\theta=1.5,~\lambda_2=0.2,~\rho=0.5,~c_1=5,$
$c_2=10$).}
\begin{center}
\begin{tabular}{|c|c|c|c|c|c|c|}\hline $\lambda_1$ &$\hat{a}_{BS}$
&$\hat{b}_{BS}$ &$\hat{a}_{BL}$ &$\hat{b}_{BL}$ &$\hat{a}_{BE}$
&$\hat{b}_{BE}$\\
\hline$\;\;0.05\;\;$ & $4.05366$ & $3.646147$ & $1.435573$ &
$0.3633854$ & $3.447897$ & $1.70214$\\
\hline $\;\;0.1\;\;$ & $2.443404$ & $3.356118$ & $0.9997084$ &
$0.368385$ & $1.866178$ & $1.464352$\\
\hline$\;\;0.2\;\;$ & $1.656511$ & $3.069585$ & $0.761078$ &
$0.3770261$ & $1.136215$ & $1.360706$\\
\hline $\;\;0.4\;\;$ & $1.156696$ & $2.816409$ & $0.5889041$ &
$0.3915007$ & $0.71128$ & $1.323718$\\
\hline$\;\;0.7\;\;$ & $0.893846$ & $2.762379$ & $0.486466$ &
$0.4145108$ & $0.5056924$ & $1.411568$\\
\hline$\;\;1.0\;\;$ & $0.7624006$ & $2.843336$ & $0.4256194$ &
$0.4462848$ & $0.4065848$ & $1.580224$\\
\hline$\;\;1.5\;\;$ & $0.6312949$ & $3.169195$ & $0.3620832$ &
$0.4954608$ & $0.3158953$ & $1.916876$\\
\hline$\;\;2.0\;\;$ & $0.535375$ & $3.588683$ & $0.3153288$ &
$0.5511547$ & $0.2584508$ & $2.311455$\\\hline
\end{tabular}
\end{center}
\end{table}

\begin{table}[h]
\caption{Table for different values of $\lambda_2$
($n=50,~r=15,~\theta=1.5,~\lambda_1=0.1,~\rho=0.5,~c_1=5,$
$c_2=10$).}
\begin{center}
\begin{tabular}{|c|c|c|c|c|c|c|}\hline $\lambda_2$ &$\hat{a}_{BS}$
&$\hat{b}_{BS}$ &$\hat{a}_{BL}$ &$\hat{b}_{BL}$ &$\hat{a}_{BE}$
&$\hat{b}_{BE}$\\
\hline$\;\;0.05\;\;$ & $3.292485$ & $10.68008$ & $0.9125832$ &
$0.5066623$ & $2.151261$ & $4.835694$\\
\hline $\;\;0.1\;\;$ & $2.926236$ & $5.977912$ & $0.9462215$&
$0.4347056$ & $2.061211$ & $2.609884$\\
\hline$\;\;0.2\;\;$ & $2.596785$ & $3.441812$ & $1.032488$ &
$0.3696412$ & $2.017182$ & $1.512799$\\
\hline $\;\;0.4\;\;$ & $2.485133$ & $1.934206$ & $1.192755$&
$0.3050227$ & $2.088596$ & $0.8874038$\\
\hline$\;\;0.7\;\;$ & $2.507843$ & $1.179365$ & $1.427955$ &
$0.2527033$ & $2.221971$ & $0.5899946$\\
\hline$\;\;1.0\;\;$ & $2.621771$ & $0.850314$ & $1.557053$ &
$0.2211348$ & $2.378278$ & $0.4535783$\\
\hline$\;\;1.5\;\;$ & $2.847125$ & $0.5858818$ & $1.835657$ &
$0.1876845$ & $2.630853$ & $0.3536515$\\
\hline$\;\;2.0\;\;$ & $3.064069$ & $0.444568$ & $2.00267$ &
$0.164858$ & $2.849913$ & $0.2923578$\\\hline
\end{tabular}
\end{center}
\end{table}

\begin{table}[h]
\caption{Table for different values of $c_1$ and $c_2$
($n=50,~r=15,~\theta=1.5,~\lambda_1=0.1,~\lambda_2=0.2,$
$\rho=0.5$).}
\begin{center}
\begin{tabular}{|c|c|c|c|c|c|c|}\hline $c_1$ &$\;\;5\;\;$
&$\;\;10\;\;$ &$\;\;20\;\;$ &$\;\;40\;\;$ &$\;\;70\;\;$
&$\;\;100\;\;$\\

\hline $\hat{a}_{BL}$ & $1.043313$ &$0.6219532$  & $0.3624302$
&$0.1955869$& $0.1200798$ &$0.1998313$ \\
\hline $c_2$ &$\;\;5\;\;$ &$\;\;10\;\;$ &$\;\;20\;\;$
&$\;\;40\;\;$ &$\;\;70\;\;$ &$\;\;100\;\;$\\
\hline$\hat{b}_{BL}$  & $0.2280337$ & $0.1650221$ & $0.1116809$
&$0.07163291$& $0.0484698$ & $0.03745599$\\\hline
\end{tabular}
\end{center}
\end{table}

\begin{table}[h]
\caption{Variance table for different values of $n$
($r=15,~\theta=1.5,~\lambda_1=0.1,~\lambda_2=0.2,~\rho=0.5,~c_1=5,$
$c_2=10$).}
\begin{center}
\begin{tabular}{|c|c|c|c|c|c|c|}\hline $n$ &$v\hat{a}_{BS}$
&$v\hat{b}_{BS}$ &$v\hat{a}_{BL}$ &$v\hat{b}_{BL}$
&$v\hat{a}_{BE}$
&$v\hat{b}_{BE}$\\
\hline$\;\;20\;\;$ & $2.786046$ & $2.801144$ & $0.1687665$ &
$0.007764694$ & $2.288183$ & $ 0.5741148$\\
\hline $\;\;30\;\;$ & $2.457465$ & $2.236923$ & $0.1677134$&
$0.004167229$ & $2.151588$ & $0.4579437$\\
\hline$\;\;50\;\;$ & $1.888847$ & $1.857697$ & $0.1675985$ &
$0.002163288$ & $1.791755$ & $0.3609909$\\
\hline $\;\;80\;\;$ & $1.476776$ & $1.527406$ & $0.1652423$&
$0.001411586$ & $1.456342$ & $0.2941667$\\
\hline$\;\;90\;\;$ & $1.435796$ & $1.490229$ & $0.1620072$ &
$0.00125574$ & $1.424549$ & $0.2921165$\\
\hline$\;\;100\;\;$ & $1.398129$ & $1.306582$ & $0.1686950$ &
$0.001065662$ & $1.401088$ & $0.2654663$\\\hline
\end{tabular}
\end{center}
\end{table}

\begin{table}[h]
\caption{Risk table for different values of $\rho$
($n=50,~r=15,~\theta=1.5,~\lambda_1=0.1,~\lambda_2=0.2,~c_1=5,$
$c_2=10$).}
\begin{center}
\begin{tabular}{|c|c|c|c|c|c|c|c|}\hline $\rho$ &$R^E_{BS}$
&$R^E_{BL}$ &$R^E_{BE}$\\
\hline$\;\;-1.0\;\;$ & $5.378928$ & $37.31374$ & $0.2712732$\\
\hline $\;\;-0.8\;\;$ & $5.012633$ & $68.0811$ & $0.2837661$\\
\hline$\;\;-0.5\;\;$ & $4.726565$ & $75.09191$ & $0.2942592$\\
\hline $\;\;-0.2\;\;$ & $4.696967$ & $27.32544$ & $0.3198501$\\
\hline $\;\;-0.1\;\;$ & $4.579706$ & $32.56944$ & $0.3070936$\\
\hline$\;\;0.0\;\;$ & $4.320078$ & $123.9033$ & $0.3094569$\\
\hline$\;\;0.1\;\;$ & $4.137500$ & $21.84179$ & $0.2806128$\\
\hline$\;\;0.2\;\;$ & $4.097972$ & $48.70084$ & $0.2945915$\\
\hline$\;\;0.5\;\;$ & $3.628694$ & $36.62481$ & $0.3146239$\\
\hline$\;\;0.8\;\;$ & $3.261719$ & $22.28436$ & $0.3165357$\\
\hline$\;\;1.0\;\;$ & $3.206366$ & $78.3708$ & $0.3034506$\\\hline
\end{tabular}
\end{center}
\end{table}

\end{document}